\documentclass[]{spie}  

\usepackage{amsmath,amsfonts,amssymb}
\usepackage{graphicx}
\usepackage[colorlinks=true, allcolors=blue]{hyperref}
\usepackage{aas_macros}

\title{The Integration and Testing Program for the Simons Observatory Large Aperture Telescope Optics Tubes}
\begin{document}

\author[a]{Kathleen Harrington}
\author[b]{Carlos Sierra}
\author[b]{Grace Chesmore}
\author[b]{Shreya Sutariya}
\author[c]{Aamir M. Ali}
\author[d,e]{Steve K. Choi }
\author[f]{Nicholas F. Cothard}
\author[g]{Simon Dicker}
\author[h]{Nicholas Galitzki}
\author[i]{Shuay-Pwu Patty Ho}
\author[g]{Anna M. Kofman}
\author[j]{Brian J. Koopman}
\author[k]{Jack Lashner}
\author[a,b,l,m]{Jeff McMahon}
\author[d,e]{Michael D. Niemack}
\author[g]{John Orlowski-Scherer}
\author[h]{Joseph Seibert}
\author[h]{Max Silva-Feaver}
\author[d]{Eve M. Vavagiakis}
\author[g,n]{Zhilei Xu}
\author[g]{Ningfeng Zhu}

\affil[a]{Department of Astronomy and Astrophysics, University of Chicago, Chicago, IL, 60637, USA}
\affil[b]{Department of Physics, University of Chicago, Chicago, IL 60637, USA}
\affil[c]{Department of Physics, University of California, Berkeley, Berkeley, CA, 94720, USA}
\affil[d]{Department of Physics, Cornell University, Ithaca, NY 14853, USA}
\affil[e]{Department of Astronomy, Cornell University, Ithaca, NY 14853, USA}
\affil[f]{Department of Applied and Engineering Physics, Cornell University, Ithaca NY, 14853, USA}
\affil[g]{Department of Physics and Astronomy, University of Pennsylvania, Philadelphia, PA, 19104, USA}
\affil[h]{Department of Physics, University of California, San Diego, La Jolla, CA 92093, USA}
\affil[i]{Department of Physics, Stanford University, Stanford, CA 94305, USA}
\affil[j]{Department of Physics, Yale University, New Haven, CT 06520, USA}
\affil[k]{Department of Physics and Astronomy, University of Southern California, Los Angeles CA, 90089, USA}
\affil[l]{Kavli Institute for Cosmological Physics, University of Chicago, Chicago, IL 60637, USA}
\affil[m]{Enrico Fermi Institute, University of Chicago, Chicago, IL 60637, USA}
\affil[n]{MIT Kavli Institute, Massachusetts Institute of Technology, Cambridge, MA, 02139, USA}

\authorinfo{Further author information: (Send correspondence to K.H: katieharrington@uchicago.edu)}

\date{Dec. 15 2020}

\maketitle

\begin{abstract}
The Simons Observatory (SO) will be a cosmic microwave background (CMB) survey experiment with three small-aperture telescopes and one large-aperture telescope, which will observe from the Atacama Desert in Chile. In total, SO will field over 60,000 transition-edge sensor (TES) bolometers in six spectral bands centered between 27 and 280 GHz in order to achieve the sensitivity necessary to measure or constrain numerous cosmological quantities, as outlined in The Simons Observatory Collaboration et al. (2019). The 6~m Large Aperture Telescope (LAT), which will target the smaller angular scales of the CMB, utilizes a cryogenic receiver (LATR) designed to house up to 13 individual optics tubes. Each optics tube is comprised of three silicon lenses, IR blocking filters, and three dual-polarization, dichroic TES detector wafers. The scientific objectives of the SO project require these optics tubes to achieve high-throughput optical performance while maintaining exquisite control of systematic effects. We describe the integration and testing program for the SO LATR optics tubes that will verify the design and assembly of the optics tubes before they are shipped to the SO site and installed in the LATR cryostat. The program includes a quick turn-around test cryostat that is used to cool single optics tubes and validate the cryogenic performance and detector readout assembly. We discuss the optical design specifications the optics tubes must meet to be deployed on sky and the suite of optical test equipment that is prepared to measure these requirements. 
\end{abstract}

\section{Introduction}

Observations of the cosmic microwave background (CMB) continue to revolutionize our understanding of the contents and evolution of the universe\cite{Plan18vi,choi2020,aiola2020,loui17,omor17,henn18,adac20,abaz16} and Simons Observatory (SO) is one of the next observatories dedicated to extending these observations to new regimes.\cite{SO_Science_2019} SO consists of one Large Aperture Telescope (LAT) optimized to target smaller angular scale fluctuations in the CMB temperature and polarization and three Small Aperture Telescopes (SATs) targeting the larger angular scales. The LAT will observe over 40\% of the sky across six frequency bands from 27 to 280~GHz, where the bands are chosen to span the Galactic foreground minima and avoid regions of high atmospheric emission. The high resolution maps made by the LAT will contain the damping tail of the CMB power spectra, enabling measurements of cosmological parameters such as the sum of the neutrino masses.\cite{SO_Science_2019} These maps will also measure the gravitational lensing of the CMB as it traverses the universe and observe the thermal and kinematic Sunyaev–Zeldovich effects from thousands of galaxy clusters; observations of the late time universe which can shed light on the current tension between early and late-time measures of the Hubble parameter.\cite{verde_2019_tension}

The LAT has two 6~m mirrors in a cross-Dragone configuration that focus light into the Large Aperture Telescope Receiver (LATR) cryostat. This 2.4~m diameter cryostat is designed to hold up to 13~optics tubes, seven of which will be populated for the initial phase of SO.\cite{Zhu_2018} Each optics tube uses three silicon lenses, anti-reflection coated with meta-material layers diced into the surfaces,\cite{Golec2020} to reimage the incoming light onto a focal plane comprised of three Universal Focalplane Modules (UFMs),\cite{healy2020} that house the detectors and coupling optics for each frequency band. Each pixel in the UFM is dichroic and dual-polarization, meaning three flavors of UFMs at low, mid, and ultra-high frequency (LF, MF, and UHF, respectively) are required for the six frequency bands used in SO. Each optics tube will contain three UFMs of the same flavor with AR coatings and low-pass edge filters matched to those particular bands. In total, the seven optics tubes in the initial phase of SO will contain nearly 32,000 transition edge sensor (TES) detectors\cite{Galitzki_2018, stevens_2020_tes} that will be read out using microwave multiplexing ($\mathrm{\mu}$Mux) SMuRF electronics\cite{Henderson_2018,SO_SMuRF_Readout}. 

The scientific goals of SO require high sensitivity maps with low levels of systematics effects. Achieving these unprecedented requirements necessitates quality control at all levels of the design and integration process. Here we discuss the integration and testing program for the SO LATR optics tubes. In Section~\ref{sec:LATRt} we describe the single optics tube test cryostat that has been developed using an SAT cryostat to decouple the optical testing from the development of the LATR cryostat and in Section~\ref{sec:requirements} we provide an overview of the identified critical metrics that require optics tube level testing before deployment to the SO site.

\section{ Optics Tube Testing Platform: LATR Tester}
\label{sec:LATRt}

The optical testing of the LATR optics tubes was decoupled from the LATR integration and testing to optimize the personnel and time resources before deployment. A fully loaded LATR cryostat is expected to take 35~days to cool down to 100~mK\cite{Coppi_2018}. A quick turn around single-tube test cryostat will perform optical validation measurements in parallel to LATR integration and allows for more and varied cool downs of optics tubes if design changes are necessary. In addition, decoupling Optics Tube testing from LATR integration means Optics Tube testing can continue after the LATR has been shipped to Chile and LAT commissioning has begun. 

\begin{figure}
    \centering
        \includegraphics[width=\textwidth]{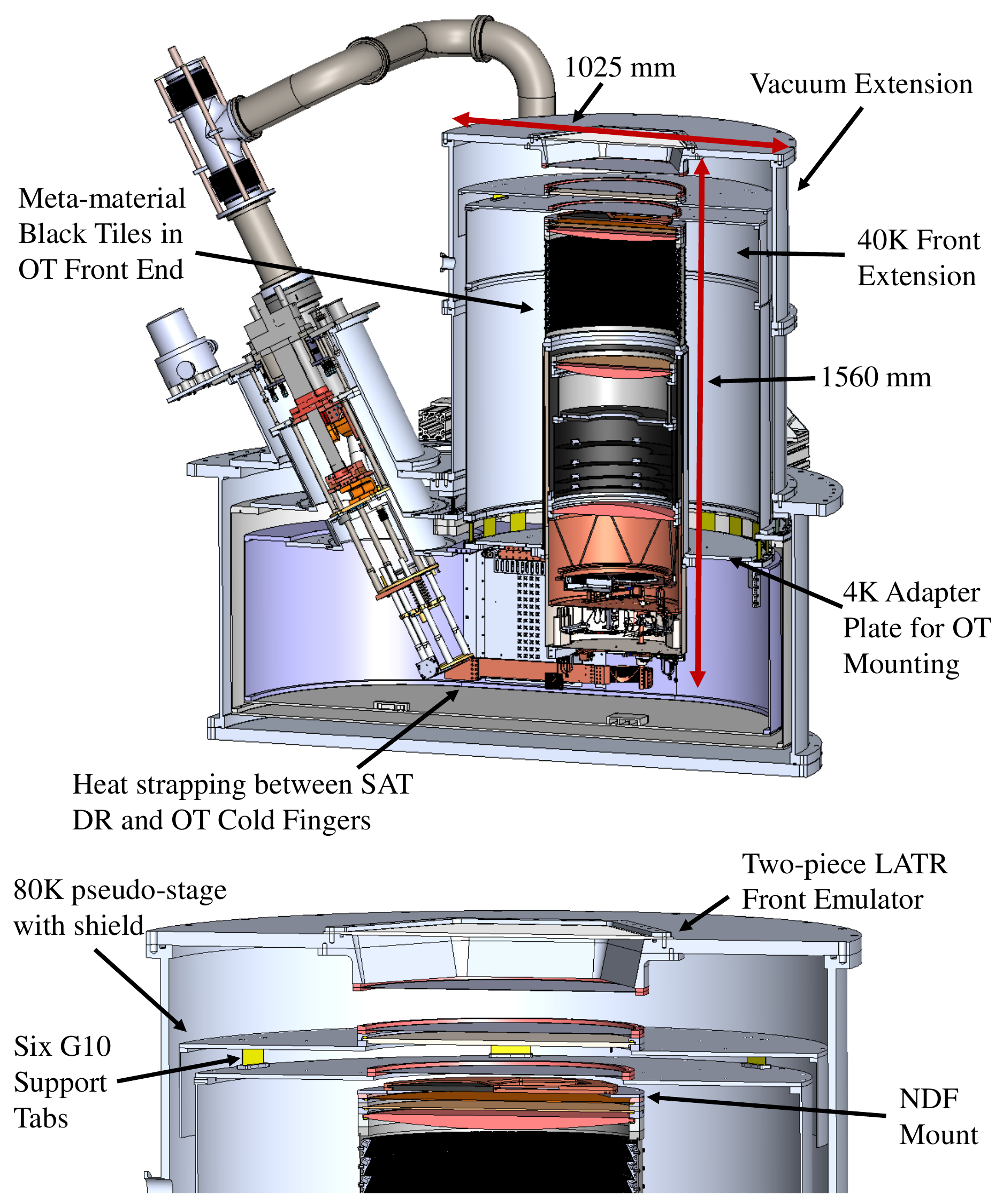}
    \caption{\label{fig:latrt_cryo} The upper figure shows a cross-section of the LATR-Tester cryostat while the lower figure shows a more detailed section of the front-end developed to emulate the optical setup of the LATR cryostat. The various alterations made to the SAT cryostat to enable LATR optics tube testing are indicated.}
\end{figure}

This LATR tester (LATRt), shown in Figure~\ref{fig:latrt_cryo}, was built out of one of the SO SAT cryostats that is slated for later deployment after LATR optics tube testing is completed. The design of the SAT cryostats was presented in Galitzki et al. 2018\cite{Galitzki_2018}. Each SAT cryostat has four temperature stages nominally at 40, 4, 1, and 0.1~Kelvin. The ``backend'' of these cryostats contains the cryocoolers and the housekeeping and detector readout paths and the ``frontend'' holds the different optical elements. Cooling power is provided by a Bluefors SD400 Dilution Refrigerator\footnote{\url{https://bluefors.com/products/sd-dilution-refrigerator/}} backed by a Cryomech PTC-420 cryocooler\footnote{\url{https://www.cryomech.com/products/pt420/}} and an additional PTC-420 mounted directly to the backend 40~K and 4~K shells. In total, each SAT is expected to have 110~W of cooling power at 40~K, 4~W at 4~K, 10~mW at 1~K, and 400~$\mathrm{\mu}$W at 100~mK. 

The mechanical design of the LATR was presented in Zhu et.~al.~2018\cite{Zhu_2018}. When compared to the SAT, the length of the LATR optics tubes is significantly longer. Converting an SAT into the LATRt required extending both the frontend vacuum shell and the 40~K frontend. This was accomplished by replacing the vacuum shell intended for the SAT half-waveplate with a longer version that placed the window the correct distance from the UFMs. The LATR window plate is about 6~cm thick, a thickness that is required due to its 2.3~m diameter. This thickness is unnecessary in the LATRt, so a second aluminum piece is attached to the vacuum side of the LATRt window plate to emulate that thickness and hold the 300~K double-sided IR (DSIR) filters\cite{cardiff} in positions to match the LATR.

In addition to the extra required length, the LATR has an addition temperature stage, nominally at 80~K, cooled by single-stage Cryomech P90 pulse-tube coolers. This stage holds additional DSIR filters and a meta-material AR coated alumina IR absorbor. Six G10 standoffs are used to build a pseudo-80~K stage in the LATRt, this stage is thermally connected to the 40~K stage via four two-layer heat-straps of 5N Aluminum. In practice, this stage is completely thermally coupled to the 40~K stage below. Lastly, a 4~K optics tube mounting plate was installed on the 4~K backend and the SAT frontend 4~K shell was removed to further emulate the LATR cryogenic setup.  

Housekeeping and detector readout is achieved with very few changes compared to the SAT and optics tube architecture. This is due to the interchangeability between the LATR and SAT systems. The warm and cold thermometry breakout boards and wiring for both cryogenic systems are nearly identical; only one additional cryogenic cable is required to incorporate LATR optics tube housekeeping into the SAT system. Similarly, the $\mathrm{\mu}$Mux readout for the LATR and SAT cryostats use identical universal readout harnesses (URHs)\cite{rao_2020} for readout components between 300~K and 4~K. Connecting the LATR optics tube cold readout assembly\cite{rao_2020} to the SAT URH only required 12 additional runs of isothermal hand-flexible coaxial cables.

\begin{figure}
    \centering
    \begin{tabular}{cc}
        \includegraphics[width=0.48\textwidth]{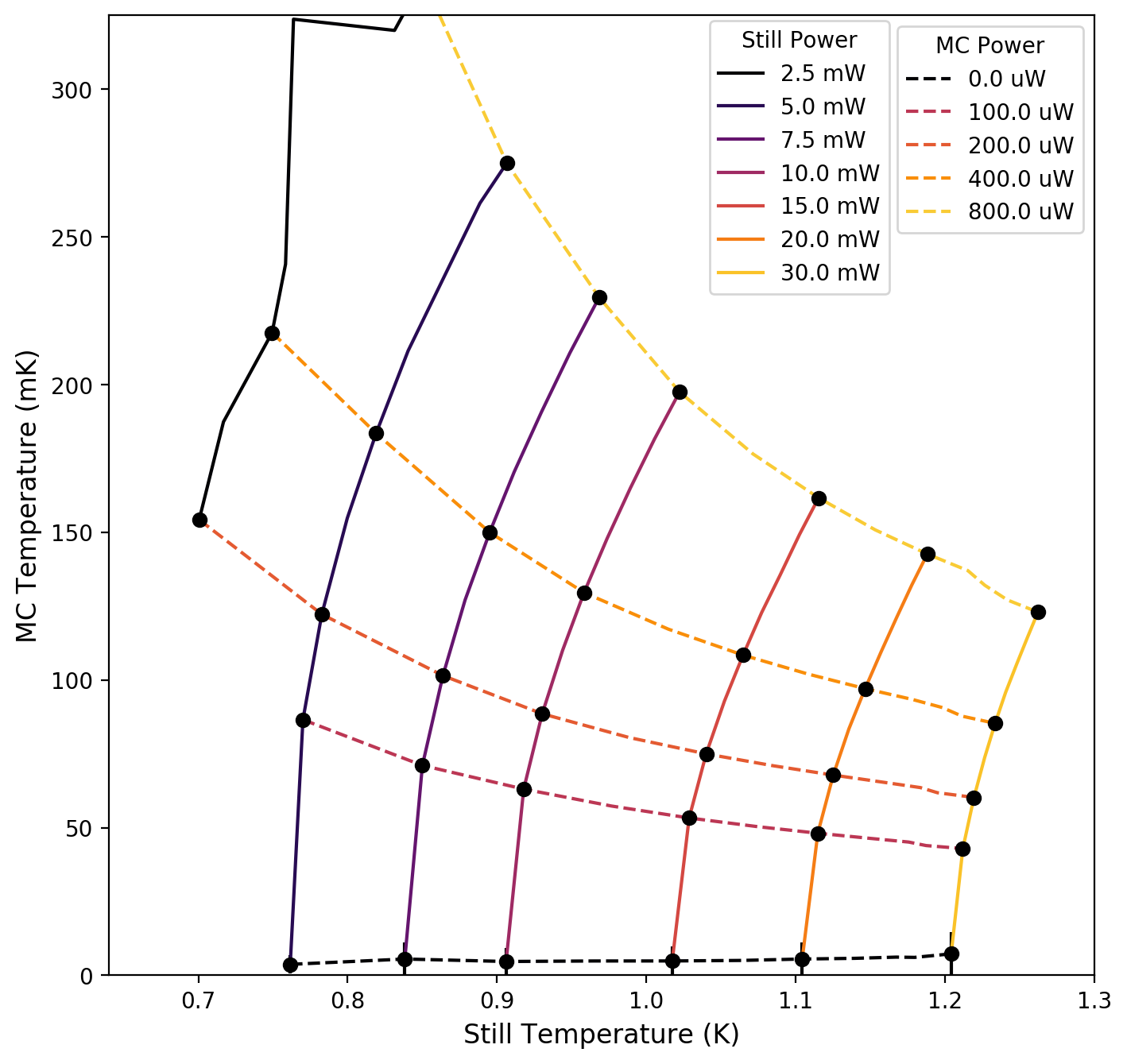} &  
        \includegraphics[width=0.48\textwidth]{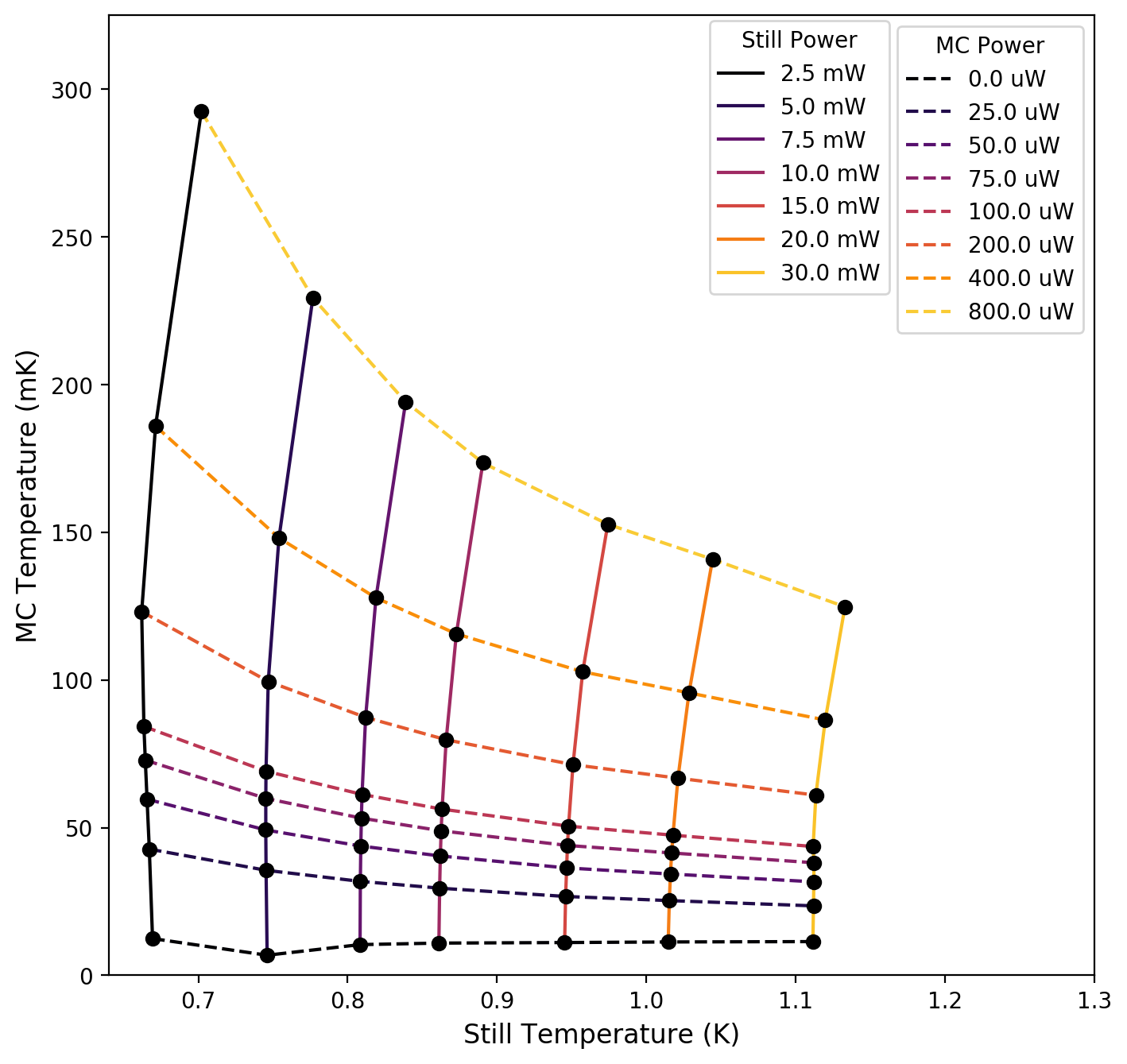} \\
    \end{tabular}
    \caption{Load curves taken with the LATRt Bluefors SD400 Dilution Refrigerator (left) held vertically before installation in the LATRt cryostat and (right) held at $27.5^\circ$ after installation in the LATRt but with no connections on the still or mixing chamber.}
    \label{fig:DR_load_curves}
\end{figure}

The LATRt completed several integration and calibration cooldowns before the first optics tube installation. The installations of the pulse-tube cryocooler and the dilution refrigerator were validated. Load curves at all stages were taken before the optics tube was installed in the LATRt. Figure~\ref{fig:DR_load_curves} shows the Bluefors SD400 DR capacity map before it was installed in the LATRt (left) and after it was installed in the LATRt (right). In the LATRt it operates at a $27.5^\circ$ angle with respect to gravity,\footnote{The tilting of the DR is due to the in-lab setup, the deployment configuration for the SAT DRs is close to $0^\circ$.} which reduces its cooling capacity. The base temperature before installation was below 7~mK and it was about 12~mK in the LATRt without an optics tube attached. Even tilted at $27.5^\circ$ and set inside a 4~K shell, the DR mixing chamber has 400~$\mathrm{\mu}$W of available cooling power at 100~mK. This level of cooling power is nearly equal to that of the LATR DR\cite{xu2020} and completely sufficient for running one LATR optics tube.


In subsequent tests, the LATRt has been used to measure the cryogenic optical loading of the front-end filtering scheme for the LATR. The combined optical loading on the 80~K and 40~K stage was 9.7~W for the set of filters tested in the LATRt.\footnote{A similar level of loading was observed for this filter set in the LATR\cite{xu2020}} This was higher than the expected loading of 3.9~W and was traced to a higher than expected blue-leak above the DSIR cutoff frequency. The optics tube and URH were also installed and the parasitic cryogenic loading from the optics tube was measured to be $4.2\pm0.2$~$\mathrm{\mu}$W on the 100~mK stage, in good agreement with expectations and measurements in the LATR. With this final set of validations complete, the LATRt is ready to proceed with the LATRt optics tube testing program.

\section{Requirements for Integration Testing}
\label{sec:requirements}

The LATRt testing program is designed to ensure the LATR Optics Tube design achieves, at a minimum, the baseline noise requirements laid out in Simons Observatory Collaboration et al.\cite{SO_Science_2019}. Discounting survey parameters, such as scan pattern and integration time, the noise levels in SO maps will be determined by the array Noise Equivalent Temperatures (NETs) of each deployed optics tube. As discussed in Hill et al.\cite{Hill_2018}, if every detector in an array has the same properties, the array NET is

\begin{equation}
    \mathrm{NET}_\mathrm{arr} = \frac{\Gamma}{\sqrt{2}} \frac{\sqrt{\mathrm{NEP}_\mathrm{ph}^2 + \mathrm{NEP}_\mathrm{g}^2 + \mathrm{NEP}_\mathrm{read}^2}}{\left( \frac{dP}{dT_\mathrm{CMB}} \right) \sqrt{Y N_\mathrm{det}}}.
    \label{eq:net_arr}
\end{equation}

\noindent In this equation, the Noise Equivalent Power (NEP)s are the contributions to the total single-detector NEP (in $\mathrm{W}/\sqrt{\mathrm{Hz}}$ units) from photons, thermal carrier noise, and readout noise, respectively, $\frac{dP}{dT_\mathrm{CMB}}$ is the end-to-end optical efficiency that is used to convert between units of power and temperature, $N_\mathrm{det}$ is the number of detectors in the array, $Y$ is the yield, the fraction of working detectors, and $\Gamma$ is the level of optical white noise correlations. The LATRt testing program focuses on the aspects of this equation that depend on the performance of the LATR Optics Tubes and cannot be tested at smaller scales (i.e. testing of individual detector modules). The testing program will validate the designs of the different flavors (LF, MF, and UHF) of optics tube, meaning one tube per flavor will undergo the tests laid out below.

\subsection{Sensitivity}

The denominator of Equation~\ref{eq:net_arr} has a few different factors that affect the overall sensitivity of the array. The $YN_\mathrm{det}$ combination represents the number of working detectors in any array. This will be measured for individual detector modules (UFMs), and again once the the UFMs have been integrated into the Optics Tube with all the associated readout components\cite{rao_2020}. Each MF and UHF UFM will contain 864~optical detectors per frequency band while each LF UFM will have 74~optical detectors per band. The SO baseline noise projections assume a 70\% yield across all arrays. 

Each Optics Tube tested in the LATRt will be cooled down with a blank-off plate and a cold load at the end of the optics tube at 4~K. IV curves, saturation powers, and dark noise levels will be measured for all detectors in this configuration to ensure the arrays have been successfully integrated into the optics tube. 

The blank-off plate will be removed and replaced with up to three different neutral density filters (NDFs) designed to cut the optical loading on the detectors to less than $1/2$ of the detector's saturation power. This will enable in-lab optical testing where the ${\sim}293$~K environment is much brighter than the ${\sim}10$~K atmospheric signal expected during normal operations at the Atacama site. The NDFs used will be primarily absorptive, made out machinable Eccosorb\footnote{https://www.laird.com/products/absorbers/\\injection-molded-machined-cast-liquids-and-microwave-absorbing-thermoplastic/eccosorb-mf} of various loss levels and thicknesses which will be tuned to the frequency bands under test. The 4~K NDF mounting plate is designed to allow changes between different NDF options in situ, without removal of the entire optics tube. Successive cool-downs with different NDF properties can be used to further calibrate the transmission of the NDFs or to optimize the optical loading between the upper and lower bands on each UFM.

The end-to-end optical efficiency, the $dP/dT_\mathrm{cmb}$ factor in the denominator of Equation~\ref{eq:net_arr}, is a component that cannot be tested without a fully integrated optics tube. As described in Hill~et.~al.\cite{Hill_2018}, the end-to-end optical efficiency accounts for the transmission through every optical element as well as the stop efficiency and the detector efficiency. This factor will be measured using chops of beam filling thermal loads ranging from 293 to 350~Kelvin. These thermal loads are 1~m by 1~m plates covered with meta-material absorbing black tiles (see Ref. ~\citenum{Xu_2020_blacktiles}) mounted on sliders positioned above the window of the cryostat. Heater elements on the back of these plates will actuate the temperatures of the tiles and an infrared thermometer will be used to measure the temperatures on the side of the plates facing the cryostat window. These measurements will be performed with multiple NDF options in front of the same detector/optical setup to simultaneously calibrate the NDF transmission and the optical efficiency of the rest of the system. The combined in-lab measurements of detector yield and end-to-end optical efficiency must be sufficient to achieve the baseline array NETs required for the SO science goals.

    
\subsection{Detector Noise Equivalent Power}

The numerator of Equation~\ref{eq:net_arr} contains the different contributions to the single-detector NEPs for the array. Simons Observatory, as with most CMB experiments, aims to be photon noise limited, meaning the $\mathrm{NEP}$ due to photons is larger than the thermal carrier noise and readout noise. For the wavelengths of interest here, both shot noise and wave noise contribute to $\mathrm{NEP}_\mathrm{ph}$, but both of these factors depend on the total optical loading on the detectors. 

Due to the nature of in-lab characterizations, including the need for NDFs and absence of the 6~meter LAT mirrors, the optical loading during optics tube testing will not match the expected on-sky loading. However, several instrument performance parameters that are expected to impact the on-sky optical loading can be verified in-lab. The two of these that are most important for the science goals of the LAT are end-to-end bandpass and spill to 300~K. 

\begin{figure}
    \centering
    \begin{tabular}{cc}
        \includegraphics[width=0.52\textwidth]{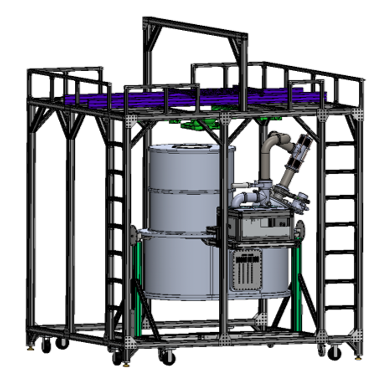} & 
        \includegraphics[width=0.44\textwidth, angle=90]{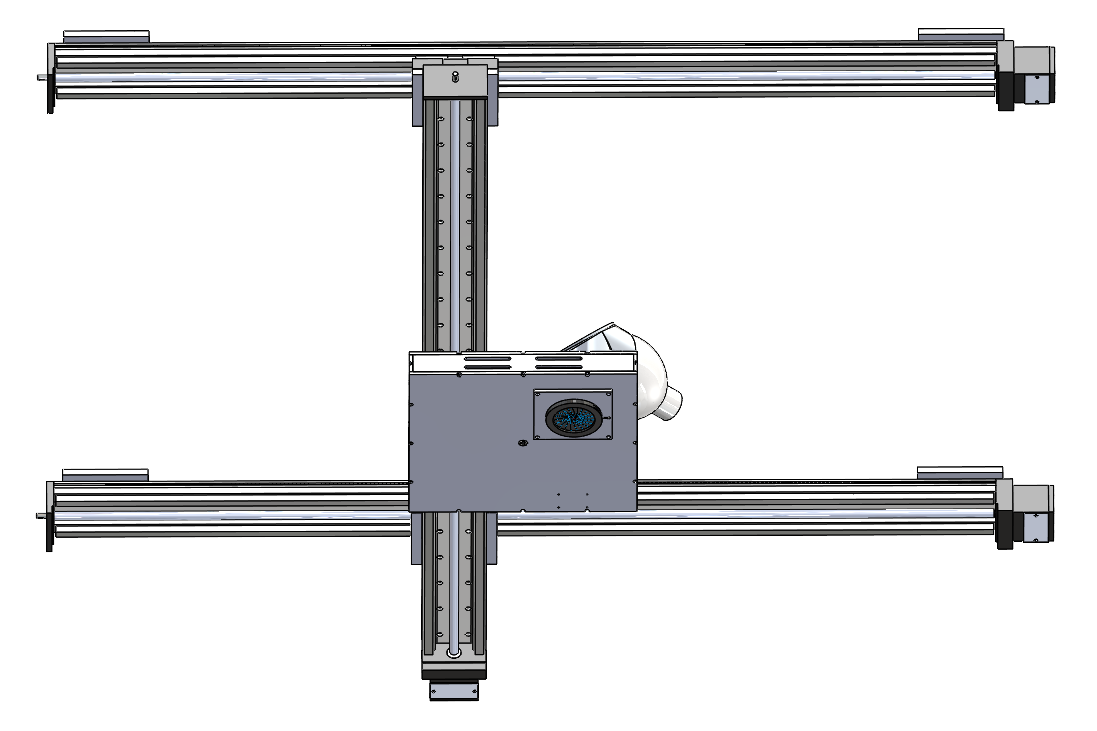} \\
    \end{tabular}
    \caption{(Left) The instrument mounting structure designed to hold the different sets of optical test equipment and roll over the LATRt cryostat during operation. Sliders, shown in blue, will be used to measure the end-to-end optical efficiency through chopping of beam-filling thermal sources at different temperatures. (Right) The XY-stage with 150~cm of travel in both directions which mounts to the underside of the instrument mounting structure and will be used for both thermal and holographic beam mapping.}
    \label{fig:test_eq}
\end{figure}
\subsubsection{Bandpass}

The end-to-end bandpasses of a CMB instrument affects both the overall sensitivity of the instrument and more systematic effects such as the separation of galactic foregrounds emission from the CMB signals. Investigating the galactic foreground separation requirements leads to a final on-sky bandpass calibration requirement of better than 0.5\%\cite{Ward_2018} for the SO instruments. However, the requirements for in-lab validation measurements are less stringent because these measurements are primarily concerned with ensuring the sensitivity of the instrument. 


A Fourier Transform Spectrometer (FTS) in the style of the PIXIE\cite{pixie_2011} instrument with refractive coupling optics will be used to measure the end-to-end bandpasses for the optics tubes. The band-edges for the integrated optics tube and UFMs must be within 2\% of the designed values, ensuring the edges are far enough from atmospheric absorption features to achieve the desired on-sky optical loading. The measured bandpasses will be propagated through the SO instrument sensitivity tracking, which must result in projected noise levels consistent with those in Simons Observatory Collaboration et al.\cite{SO_Science_2019}. High precision bandpass calibrations will be performed on-site where NDFs are unnecessary. 


\subsubsection{Spill to 300~K}

The fraction of the beam that spills to 300~Kelvin has a significant effect on the optical loading on the detectors. During the design of the LAT instrument, it was found that spill to 300~K was one of the main drivers in the projected sensitivity overall and that reducing this spill could significantly reduce the noise in the resulting SO maps. 

As described in Gudmundsson et al.\cite{Gudmundsson_2020}, a substantial effort has gone into controlling the scattering inside the optics tubes to reduce this warm spill, including the design of meta-material absorbing black tiles\cite{Zhu_2018} that will be used instead of baffles blackened with carbon-loaded Stycast\footnote{\url{https://www.henkel-adhesives.com/us/en/product/loctite_stycast_2850ft.html}} in the upper region of the optics tubes. The LATRt will be verifying that these efforts were successful before the optics tubes are deployed to the site. 

The upgraded optics tube baffling design reduces the expected scattering at large angles ($\gtrsim30^\circ$) from -$25$~dB to -$50$~dB levels. Similarly, constraining the spill outside of the secondary mirror to a 1\% level will require mapping the beam down to about -$33$~dB to measure a significant enough fraction of the total beam.

The LATRt will use two different types of beam mapping to verify the efficacy of the black tiles and constrain the expected warm spill. First, we use a $600^\circ$C IR source mounted behind a spinning chopper wheel with an adjustable aperture. The source will be mounted on an XY~stage with 150~cm of travel in both directions. With this source, a 25~mm aperture, and a 5\% transmissive NDF, we expect to require 1~second of integration time to measure a -36~dB signal at a signal-to-noise ratio of one. This sensitivity is expected to be sufficient to verify that the black tiles perform better than the carbon-loaded stycast simulations and to map the beam to low enough levels to constrain the warm spill to a 1\% level. 

Measuring the amount of scattering expected from the black tiled optics tube requires measuring the beams down to a -50 to -60 dB level. This is unlikely to be possible with thermal sources in lab but may be possible with holographic measurements of the detector beams. This measurement uses a coherent source and receiver to measure the amplitude and phase of the electric fields across a two-dimensional surface in front of the optics tube window. A coherent receiver will be mounted on a feedhorn array at the optics tube focal plane, with signals transmitted through the cryogenic coaxial cables used for the SMuRF $\mathrm{\mu}$Mux system. The source will be mounted on the same XY~stage used for the thermal beam mapping. These beam maps will be compared to simulations of the expected beams at the measurement plane to verify the effectiveness of the black tiles. 

    

\begin{figure}
    \centering
    \includegraphics[width=\textwidth]{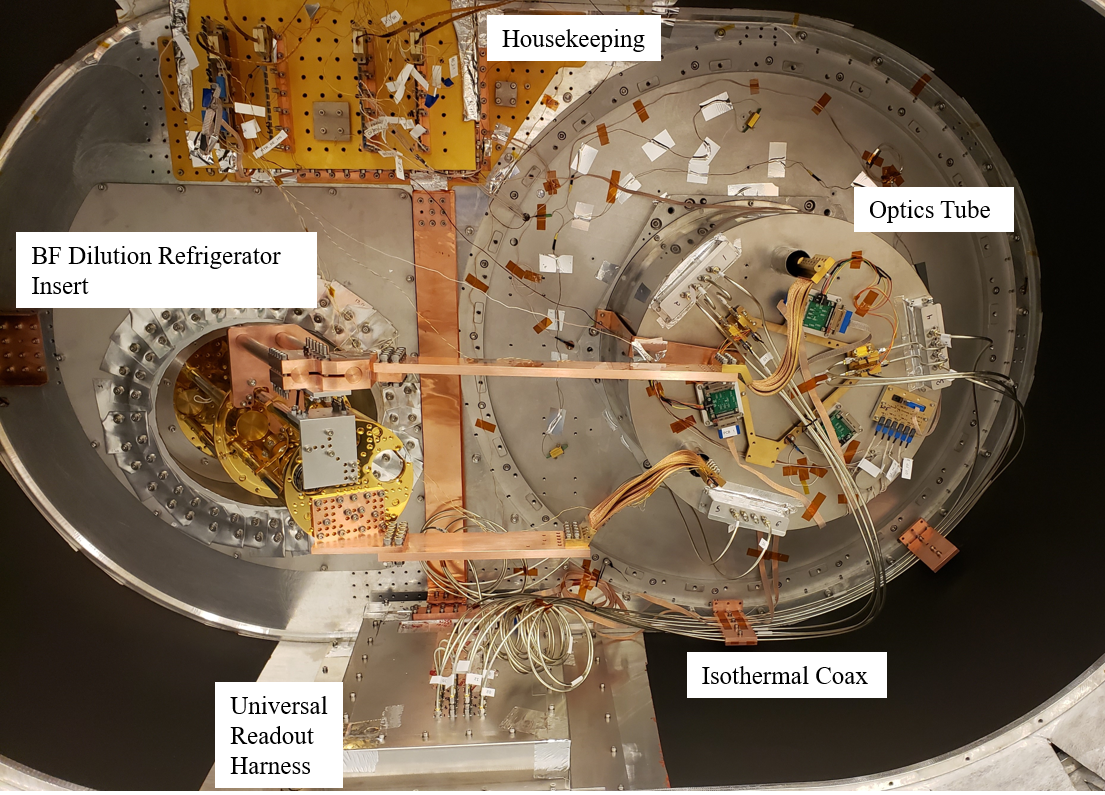}
    \caption{The backend of the integrated LATRt cryostat, showing the installed optics tube, universal readout harness, housekeeping, and Bluefors DR insert. Custom heat straps are used to connect the DR stages to the optics tube cold fingers that provided a cooling path the the 1~K and 100~mK stages.}
    \label{fig:latrt_backend}
\end{figure}
\section{Status and Outlook}

At the time of these proceedings, the LATRt cryostat, pictured in Figure~\ref{fig:latrt_backend}, has been integrated and cryogenically tested with all major components except UFMs (detectors) and the optics tube IR filters. Those will be installed next and the LATRt Testing program will begin characterizing the performance of the first MF~optics tube. In depth testing is currently planned for one of each flavor (LF, MF, UHF) of optics tube to verify the design of each of type of optics tube. Testing on the other four optics tubes that are part of the nominal-SO plan will depend on hardware availability and deployment schedules.

\acknowledgments 
 
This work was funded by the Simons Foundation (Award \#457687, B.K.). KH was supported by the U.S. Department of Energy, Office of Science, under Award Number DE-SC0015799. ZX is supported by the Gordon and Betty Moore Foundation.

\bibliography{sources}
\bibliographystyle{spiebib}

\end{document}